# Binding Energies of the Deuteron, the Neutron and the Alpha Particle From a Theoretical Geometric Model.

Gustavo R. González-Martín[a]

**Departamento de Física, Universidad Simón Bolívar, Apartado 89000, Caracas 1080-A, Venezuela.**

We assume a triple geometric structure for the electromagnetic nuclear interaction. This nuclear electromagnetism is used to calculate the binding energies of the deuteron and the neutron. The corresponding Pauli quantum wave equation in a geometric theory, with the SU(2) electromagnetic coupling instead of the standard "minimal" coupling, contains a $1/r^4$ short range attractive magnetic potential term. This term, produced by the odd part of the electromagnetic potential, may be responsible for a strong nuclear interaction. An approximation for the resultant wave equation leads to the modified Mathieu equation for the radial eigenvalue equation. Completely theoretical calculations give *2.205 Mev., 0.782 Mev.* and *27.6 Mev.* for the binding energies of the deuteron, the neutron and the alpha particle respectively. These values admit correction factors due to the approximations made.

21.10Dr, 21.60.-n, 14.20.Dh, 02.20.Qs

[a] Webpage http://prof.usb.ve/ggonzalm

# 1- Introduction.

A modified Pauli equation[1, 2], where a non standard electromagnetic coupling introduces extra minus signs due to its SU(2) structure, has been discussed. This equation was useful in understanding the proton and neutron magnetic moments[2]. The relevance of the presence of these extra signs was overlooked for many years. For certain applications it is sufficient to only use a non standard triple electromagnetic $SU(2)_Q$ coupling in the standard Pauli equation[3]. Electromagnetism is represented by a connection potential corresponding to the $SU(2)_Q$ generators which are subject to quantization in the same manner as the angular momentum generators. In particular, this geometric coupling provides an attractive $1/r^4$ magnetic potential that may be important in nuclear processes. Interest in the effects of electromagnetism on the nuclear structure is not new.[4] Nevertheless, here we present new effects due to its possible nuclear electomagnetic triple structure.

# 2- The modified Pauli equation.

The modified Pauli equation may be expressed using the main involution of the Minkowski space Clifford algebra applied to its electromagnetic su(2) subalgebra. In geometric units $c = h = e = 1$, the equation is[2]

$$i\nabla_0 \psi = \left[ \frac{\left(i\nabla + \left(^+A - ^-A\right)\right)\cdot\left(i\nabla + \left(^+A + ^-A\right)\right)}{2m} - \left(^+A_0 + ^-A_0\right) - \frac{\sigma\cdot\nabla\times\left(^+A + ^-A\right)}{2m} + \frac{\sigma\cdot ^-A\times\nabla}{m} \right]\psi \quad (1)$$

in terms of the even and odd parts of the 3-vector potential $A$ and the scalar potential $A_0$. The absolute value of the charge $e$ is incorporated in the su(2) connection or 4-vector potential $A$. The wave function is an eigenvector of the $SU(2)_Q$ physical charge operator with charge eigenvalues corresponding to the representation. The 4-potential $A_\mu$ has an orientation in the Clifford algebra with a geometric algebra polar angle $\Theta'$ with respect to the standard electromagnetic potential, required by quantization[2] of the $SU(2)_Q$ generator in the $(n,i)$ representation as indicated in the figure and the appendix in terms of a quaternion base $q^a$. The angle $\Theta_{½}$ of the fundamental (½,½) representation is $\pi/3$, approximately equal to the complement of Weinberg's angle and may suggest its geometric interpretation. When the angle $\Theta$ is written without indices it should be understood that it refers to the fundamental representation. At very small nuclear distances the action of potentials $^-A_\mu$ or $A_\mu$ on a common eigenvector can be simply expressed in terms of the action of the even potential $^+A_\mu$,

$$A_\mu \psi = \left(^+A_\mu + ^-A_\mu\right)\psi = \left(n + \lambda(i)\right)\psi_\mu = \left(1 + \frac{\lambda(i)}{n}\right)n\psi_\mu = \left(1 + \tan\Theta_n^i\right)^+A_\mu \psi \quad . \quad (2)$$

In this manner we obtain for very small nuclear distances

$$[-\nabla^2 + 2i\,^+A\cdot\nabla - \left(\tan^2\Theta - 1\right)^+A^2 - \left(1 + \tan\Theta\right)\sigma\cdot\nabla\times\,^+A + 2\tan\Theta\left(\sigma\cdot\,^+A\times\nabla\right)$$
$$- 2m\left(1 + \tan\Theta\right)^+A_0]\psi = 2mE\psi \quad . \quad (3)$$

The $SU(2)_Q$ connection is a short range field.[5] At large atomic distances, $\Theta'$ is zero and we get the standard Pauli equation for the standard U(1) electromagnetic even potential $^+A$,

$$[-\nabla^2 + 2ie\,^+A\cdot\nabla + e^2\,^+A^2 - e\sigma\cdot\nabla\times\,^+A - 2me\,^+A_0]\psi = 2mE\psi \quad . \quad (4)$$

The tangent of the geometric angle $\Theta$ in eq. 3 is $3^{½}$,

$$\left[-\nabla^2 + 2i\,^+A\cdot\nabla - 2\,^+A^2 - \left(1 + \sqrt{3}\right)\sigma\cdot\nabla\times\,^+A + 2\sqrt{3}\left(\sigma\cdot\,^+A\times\nabla\right) - 2m\left(1 + \sqrt{3}\right)A_0\right]\psi = 2mE\psi .$$

(5)

In terms of the magnetic moment $\mu$ and the charge $q$ of a point source, the equation is

$$\left[-\nabla^2 + 2i\frac{\vec{\mu}\times\vec{r}}{|r|^3}\cdot\vec{\nabla} - 2\frac{(\vec{\mu}\times\vec{r})^2}{|r|^6} - (1+\sqrt{3})\vec{\sigma}\cdot\vec{B} + 2\sqrt{3}\left(\vec{\sigma}\cdot\frac{\vec{\mu}\times\vec{r}}{|r|^3}\times\vec{\nabla}\right) - \frac{2m(1+\sqrt{3})q}{r}\right]\psi = 2mE\psi \quad (6)$$

where we see that the dominant effects are the magnetic terms which include an attractive $r^{-4}$ potential.

It should be indicated that the gyromagnetic factor $g$ is an internal property of a particle or excitation which is determined by the geometric (quantum) electromagnetic operators $A$ present in the internal magnetic energy term of its internal motion equation. Therefore, it is convenient to recognize a pure geometric factor $\gamma$ in the magnetic moment $\mu$ which includes these factors but excludes the mass which represents a secondary inertial reaction in the charge motion. Depending on the particular system considered this mass may correspond to the system reduced mass or to a total mass.

## 3- The proton-electron-proton model for the deuteron.

Now consider that the equation represents a system of one electron and a few protons moving about the system center of mass. The fields are dominated by the electron magnetic field because its higher magnetic moment $\mu$. In general, we consider that the resultant SU(2) field is characterized by the system reduced mass. The system reduced mass $m$ is essentially the electron mass. In particular, we shall consider a naive model[b] for the deuteron as an 4-excitation $(p,p,e,\bar{\nu})$ of a nonlinear system where $(e,\bar{\nu}) \equiv e'$ is a spin 1 electronic excitation subsystem which has the charge, mass and magnetic moment of the electron and generates the sole magnetic field $B_e$ centered close to the 4-excitation system center of mass. The electrostatic repulsion between the protons is dominated by the presence of the negative electron charge close to the center of mass of the protons. The operators $^+A_e$ of the electron $e$ and $A_p$ of the protons $p$ in the stable system $(p,p,e')$ constitute a single operator $A$, which depends on the reduced mass, and determines the fundamental SU(2) excitation. In other words, the resultant 4-potencial $A$ has the Clifford algebra orientation required by the fundamental $SU(2)_Q$ quantum representation of the generator $A$. This determines the internal geometric polar angle $\Theta$ of $A$ with respect to the standard electromagnetic potential $^+A$. The resultant field $A$ affects all system components and in particular the protons. The reduced mass is the electron mass $m$ and the internal interaction energy proper value is $E$. The dominant internal magnetostatic energy should be the one that holds the 4-excitation system together. We may say that this is a symmetric quasi-static disposition of the two protons around the excited electron in a proton-electron-proton formation that facilitates a stable solution[c].

Because of the symmetry we may use cylindrical coordinates with the $z$ axis along the electron magnetic moment $\mu$. This equation is not separable[6] due to the external spatial polar angle $\theta$ dependence contained in the cross products,

$$\left[\begin{array}{l}\nabla^2 + 2m^2\varepsilon^2 + \dfrac{2\mu^2}{\rho^4\left(1+\dfrac{z^2}{\rho^2}\right)^3} - \dfrac{(1+\sqrt{3})\mu}{\rho^3\left(1+\dfrac{z^2}{\rho^2}\right)^{5/2}}\left(2\sigma^3\left(1-\dfrac{z^2}{2\rho^2}\right) - 3\sigma^1\dfrac{z}{\rho}\right) - \\ \\ \dfrac{2i\mu}{\rho^3\left(1+\dfrac{z^2}{\rho^2}\right)^{3/2}}\dfrac{\partial}{\partial\varphi} + \dfrac{2\sqrt{3}\mu}{\rho^2\left(1+\dfrac{z^2}{\rho^2}\right)^{3/2}}\left(\sigma^3\dfrac{\partial}{\partial\rho} - \sigma^1\dfrac{\partial}{\partial z}\right) + \dfrac{2m(1+\sqrt{3})}{\rho}\end{array}\right]\psi = 0 . \quad (7)$$

where the energy is expressed by $E = m\varepsilon^2$ as a mass fraction.

Nevertheless, since orbital and spin angular momenta tend to align with the magnetic $B$ field we should expect that the

---

b In the spirit of nuclear models where the shell model coexists with the liquid drop model, the collective model, etc.

c A similar model for the neutron is unrealistic because the analysis of the neutron highly nonlinear system corresponds to a strongly-dynamic asymmetric disposition that would not hold the 3-excitation system $(p,e,)$ together without collapsing.

wave function concentrates around the equatorial plane and most of the energy is in the equatorial region. A calculation of the negative gradient of the dominant potential term confirms this consideration,

$$-\nabla\left(\frac{-2\mu^2\rho^2}{\left(\rho^2+z^2\right)^3}\right) = \frac{2\mu^2}{\rho^4\left(1+z^2/\rho^2\right)^3}\left[\left(\frac{-4}{\rho}+\frac{6z^2}{\rho\left(\rho^2+z^2\right)}\right)u_\rho - \frac{6z}{\rho^2+z^2}u_z\right]. \tag{8}$$

We see that the negative gradient has direction toward the equatorial plane $z=0$ for all values of $z$ and toward the origin in an equatorial conical region defined by $z^2 < 2\rho^2$. This equatorial region is limited by two cones near the equator, characterized by an equatorial angle or latitude $\theta_e$

$$\left(\pi/2-\theta_e\right)\leq\theta\leq\left(\pi/2+\theta_e\right), \qquad \theta_e \approx z/\rho\sqrt{2}. \tag{9}$$

In the limit of very small $\rho$ the problem reduces to a bidimensional problem on the equatorial plane. Therefore, near this limit we should simplify by setting $\sin\theta \approx 1$, $z\approx 0$. Now it is convenient to approximate the potentials to very small $\rho$ by only keeping the dominant $\rho^{-4}$ potential terms, obtaining a separable approximate equation

$$\left[\nabla^2 + 2m^2\varepsilon^2 + \frac{2\mu^2}{\rho^4}\right]\psi = 0 \tag{10}$$

$$\left[\frac{\partial^2}{\partial\rho^2} + \frac{1}{\rho}\frac{\partial}{\partial\rho} + \frac{1}{\rho^2}\frac{\partial^2}{\partial\varphi^2} + \frac{\partial^2}{\partial z^2} + 2m^2\varepsilon^2 + \frac{2\mu^2}{\rho^4}\right]\psi = 0 \tag{11}$$

Since the $z$ dependence is originally through the polar angle $\theta$, it is convenient to introduce in its place, in the equatorial zone, an angular latitude coordinate $\zeta$,

$$\zeta = \tan\frac{z}{\rho} \approx \frac{z}{\rho}. \tag{12}$$

In terms of this coordinate the equation becomes

$$\left[\rho^2\frac{\partial^2}{\partial\rho^2} + \rho\frac{\partial}{\partial\rho} + \frac{\partial^2}{\partial\varphi^2} + \frac{\partial^2}{\partial\zeta^2} + 2m^2\varepsilon^2\rho^2 + \frac{2\mu^2}{\rho^2}\right]\psi = 0 \tag{13}$$

because

$$\frac{\partial^2}{\partial z^2} = \frac{\partial\zeta}{\partial z}\frac{\partial}{\partial\zeta}\left(\frac{\partial\zeta}{\partial z}\frac{\partial}{\partial\zeta}\right) = \left(\frac{\partial\zeta}{\partial z}\right)^2\frac{\partial^2}{\partial\zeta^2} = \frac{1}{\rho^2}\frac{\partial^2}{\partial\zeta^2}. \tag{14}$$

We assume a separable solution in the equatorial region, of the form

$$\psi = R(\rho)\Lambda^\pm(\varphi)Z(\zeta). \tag{15}$$

The separated equation for $Z$ is

$$\frac{\partial^2 Z}{\partial\zeta^2} = \alpha^2 Z \tag{16}$$

where $\alpha$ is a real approximate separation constant.

The cylindrical symmetry implies the conservation of the angular momentum azimuthal component. The Pauli spinor $\psi$, or large component of the Dirac spinor, is a common eigenspinor of the generators of rotation $L_z$, spin $S_z$ and total rotations $J_z$, with the eigenvalues,

$$J_z \left|\lambda, \pm \tfrac{1}{2}\right\rangle = (L_z + S_z)\left|\lambda, \pm \tfrac{1}{2}\right\rangle = \left(\lambda \pm \tfrac{1}{2}\right)\left|\lambda, \pm \tfrac{1}{2}\right\rangle. \qquad (17)$$

The form of the eigenfunctions may be constructed using the exponential functions. The two spinor states may be represented by exponentiation of the spin z-axis component generator $i/2$ using as parameter the azimuthal coordinate $\varphi$ extended to spinor space. We define the spinor eigenfunction $\Lambda^\pm$ as follows,

$$\Lambda^\pm \equiv \left|\lambda, \pm \tfrac{1}{2}\right\rangle = \Lambda(\varphi)\exp\left(\tfrac{i\varphi}{2}\right) = e^{i\lambda\varphi} e^{i\varphi/2} = e^{i(\lambda \pm 1/2)\varphi} \ . \qquad (18)$$

We require that the spatial part $\Lambda$ of the eigenfunction be single valued, which determines that $\lambda$ is an integer. With this understanding, the separated equation for the angular function $\Lambda^\pm$ is

$$\frac{\partial^2}{\partial \varphi^2}\left(e^{i(\lambda \pm 1/2)\varphi}\right) = -J_z^2 \left|\lambda, \pm \tfrac{1}{2}\right\rangle = -\left(\lambda \pm \tfrac{1}{2}\right)^2 e^{i(\lambda \pm 1/2)\varphi} = -\nu^2 e^{i\nu\varphi} \ , \qquad (19)$$

where we defined the half integer quantum number $\nu$.

Values of $\lambda$ mean that the system has orbital angular momentum around its center of mass. From a mechanical point of view, this implies a rotation of the lighter electron $e$ around the center of mass of the two protons $p$. This should be inconsistent with the quasi-static mechanism, indicating an instability, and we may rule out all values $\lambda \neq 0$. The stable model might only be possible for the value $\lambda = 0$ which implies a static system.

In accordance with the indication in the first section, the magnetic moment is determined by a geometric factor $\gamma$ that depends on the fields and charges in motion. The magnetic moment is inversely proportional to the mass, which tends to oppose the motion. It is convenient to make a change of variables to a radial dimensionless complex variable $z$ rationalized to the mass

$$\rho^2 = \frac{\mu}{m\varepsilon}z^2 = \frac{\gamma}{m^2\varepsilon}z^2 \qquad (20)$$

where it is known that the electron $\gamma$ is ½. We obtain the separated radial equation which has irregular singular points at zero and infinity.[6]

$$z^2 R'' + z R' + \left[2\gamma\varepsilon\left(z^2 + \frac{1}{z^2}\right) - (\nu^2 - \alpha^2)\right]R = 0 \qquad (21)$$

and let $z = e^u$,

$$R'' + \left[2\gamma\varepsilon\left(e^{2u} + e^{-2u}\right) - (\nu^2 - \alpha^2)\right]R = 0 \ . \qquad (22)$$

The resultant equation is the modified Mathieu equation[7, 8]

$$R'' + (2q\cosh 2u - a)R = 0 \ , \qquad (23)$$

with the parameters

$$q = 2\gamma\varepsilon \ , \qquad (24)$$

$$a = \nu^2 - \alpha^2 \ , \qquad (25)$$

In the last equation the parameter $a$, which does not correspond to the $\rho^{-4}$ dominant term, should be considered an effective parameter because, in addition to $\alpha^2$, it may include contributions from its $\rho^{-3}$ associated term in equation (7). For example, if the radial coordinate $\rho$ fluctuates around an average value, it may contribute to give an effective value of unit order to $a$. This approximate result regarding the application of the Mathieu equation for the system under study might also be obtained using spherical coordinates.

## 4-Binding energy for the deuteron.

According to Floquet's theorem,[9] a solution of eq. (23) includes an overall factor $e^{su}$ where $s$ is a complex constant. If $s$ is an integer we get the Mathieu functions of order $s$. A Mathieu function is one of the few special functions which is not a special case of the hypergeometric function and therefore the determination of its proper values differs. The Mathieu functions can be expressed as a Fourier series. The series expansion coefficients are proportional to the characteristic root values $a_r$ of a set of continued fraction equations obtained from the three-term recursion relations among the expansion coefficients. These characteristic roots are obtained as a power series in $q$.

A solution of the Mathieu equation can also be expressed in terms of a series of Bessel functions with the same expansion coefficients of the Fourier series. This Bessel series is convenient to find the asymptotic behavior of the solution. By replacing the Bessel functions with the Neumann functions we obtain a solution of the second kind. Solutions of the third and fourth kind are obtained by combining the Bessel and Neumann functions to form Hankel functions. We want a regular solution that vanishes at infinity. We require that $s=0$ to avoid a singular or oscillatory asymptotic behaviour. The appropriate negative exponential behavior at infinity is provided by the radial Mathieu function of zero order, of the third kind, or Mathieu-Hankel function, indicated by $Mc_0^{(3)}(q,u)$ or $He_0(q,u)$.

It is known that the characteristic roots and the corresponding series expansion coefficients for the Mathieu functions have branch cut singularities on the imaginary $q$ axis[8, 10, 11] and the complex plane $a$ of the corresponding Riemann surfaces. Therefore, the expansion coefficients are multiple-valued. In the standard radial equation eigenvalue problem, the energy is determined by eliminating the singularities of the hypergeometric series by requiring its coefficients to vanish after a certain order. In our case, the energy should be determined by eliminating the branch cut singularities of the Mathieu series by choosing appropriate coefficients. We should require that the solution coefficients of the $R$ function be single valued as we required for the $\Lambda$ function. To eliminate the possibility of multiple-valued coefficients and obtain a regular solution we must disregard all points $q$ on all branches of the Riemann surface except the points $q_0$ common to all branches. We require that $q$ be the two common points $q_0$ on the Riemann surface branches corresponding to the first two characteristic roots $a_0$ and $a_2$ which determine the rest of the expansion coefficients of the even Mathieu functions of period $\pi$.

The values of the constant $q_0$ has been calculated[8, 10, 12] with many significant digits. The parameter $q$ determines a single degenerate negative energy eigenvalue corresponding to the states of the solution, as follows,

$$q = 2\gamma\varepsilon = \varepsilon = \pm iq_0 \qquad (26)$$

and we have a stable solution with the value of binding energy per state

$$E_0 = \varepsilon^2 m_e = -q_0^2 m_e = -(1.4687686)^2 m_e = -2.1572812 m_e \approx -1.10237 \text{ Mev.} \qquad (27)$$

According to the model, both spin $\nu = \pm \tfrac{1}{2}$ states are occupied and the total energy is

$$E = -2.20474 \text{ Mev.} \approx -2.2246 \text{ Mev.} = -U_d \qquad (28)$$

where $U_d$ is the deuteron binding energy. The total energy $E$ is the necessary energy to destroy the symmetric quasi-static arrangement.

Since the protons are in opposite angular momentum states, the deuteron spin is equal to the $e'$ system spin. If we find the correct equation for the, spin 1, $e'$ system it would display a magnetic energy term similar to the ones found previously[2] where the coefficient of the $B$ field is

$$\mu_{e'} = g_{e'}\mu_B S_{e'} = \mu_e = \pm\mu_B \qquad (29)$$

in terms of the Bohr magneton and a spin 1 matrix $S$. This equation rules out the spin 0 state for $e'$ and for the deuteron. The deuteron can only be in the $\pm 1$ spin states. It also shows that $e'$ does not have an anomalous gyromagnetic factor.

## 5- The electron-proton model for the neutron.

If only one state is occupied we could expect an unstable model for the neutron. Comparing the 1-proton and the bound deuteron 2-proton system equations, we consider that the average of the dominant electron magnetic fields $B_e$ in the corresponding equations should remain approximately equal at the respective proton positions. If the stable 2-proton system suddenly looses one proton, by a collision for example, we may assume that the potential terms, produced by the electron, remain constant for the duration of a short transition, while the system becomes unstable. The symmetric quasi-static disposition of the two protons around the electron in the proton-electron-proton formation is destroyed, leading to a strongly-dynamic asymmetric disposition in an electron-proton system. When the system looses a proton, the motion instability breaks $^+A_e$ away from $^+A_p$ in such a way that their combination does not constitute the fundamental SU(2) representation. In its place there form separate electron U(1) and proton SU(2) representations respectively. In other words, this combination now implies that the sum of the respective eigenvalues determine the neutron anomalous gyromagnetic factor[2] and its energy proper value. During this transition, the parameters associated to the magnetic properties of the wavefunction must adjust themselves to the new situation in the unstable 1-proton system. One of these parameters is the anomalous gyromagnetic factor $g$ that we recognize present in one of the potential terms as a coefficient of the $B$ field

$$2\left(1+\sqrt{3}\right)\frac{\mu}{\rho^3} = g_p \frac{\mu}{\rho^3} \ . \tag{30}$$

We now understand that the gyromagnetic factor $g$, as also indicated in the previous work,[2] is determined by the SU(2) interaction. The other parameter is the energy eigenvalue $E$. The only potential terms in equation (7) that should undergo changes are those depending on these magnetic parameters. Thus, we concentrate on the expression

$$W = 2m^2\varepsilon^2 + 2\left(1+\sqrt{3}\right)\frac{\mu}{\rho^3} = 2mE + g\frac{\mu}{\rho^3} \tag{31}$$

which is the dominant expression containing these parameters, while keeping $\rho$ and the other potential terms constant in equation (7). Therefore the total value of $W$ also remains constant while $E$ and $g$ change. Then, during the transition,

$$dW = 2mdE + \frac{\mu}{\rho^3}dg = 0 \ . \tag{32}$$

Therefore, the last equation relates the change of the energy eigenvalue to the change of the gyromagnetic factor while the su(2) magnetic fields are rearranged,

$$\frac{dE}{dg} = -\frac{\mu}{m\rho^3} \equiv K \ , \tag{33}$$

where $K$ is taken as constant. We assume that the integration constant is equal to zero, which means that $E$ and $g$ proportionally evolve toward the unstable state and $W$ is zero after the collision and during the evolution toward the unstable state. The energy eigenvalue is proportional to the gyromagnetic factor and to the intrinsic magnetic moment potential term which is a fraction of total effective potential

$$\frac{E}{g} = \frac{E_n}{g_n} = \frac{E_0}{g_p} \ . \tag{34}$$

The ratio of the nucleon energy eigenvalues is then approximately equal to the ratio of the gyromagnetic factors. We know from previous work [2] that both anomalous proton and neutron magnetic moments are determined and theoretically calculated from two different su(2) magnetic field arrangements in nucleon excitation systems with mass $m_p$. The proton-electron system undergoes su(2) magnetic field changes toward an unstable neutron system that eventually decays. The energy of the latter system is then theoretically determined by these gyromagnetic factor theoretical values

$$E_n = \frac{g_n}{g_p}E_0 = \frac{-1.967}{2.780}(-1.10237) = 0.780 \text{ Mev.} \approx m_n - m_p - m_e = 0.782 \text{ Mev.} \tag{35}$$

This system has a mass-excess equal to $E_n$ which is given-up as energy when it decays into its constituents proton and electron. Then the energy $E$ in eq.(28 ) should be interpreted as the deuteron binding energy.

The Pauli equation (6) for the internal motion relative to the center of mass, with reduced mass $m$, under the internal magnetic field, serves to calculate gyromagnetic effects and binding energies. For the neutron we have

$$\left[-\nabla^2 - (-2)\left(1+\sqrt{3}\right)\frac{g_n}{g_p}\frac{\sigma \cdot B}{2}\right]\psi = 2mE_n\psi \ . \tag{36}$$

The Pauli equation also serves to study the motion of the neutron system center of mass under an external magnetic field, now with the resultant total mass,

$$M'_n = M_p + m_e + E_n \ . \tag{37}$$

In this case the equation, with the same geometric gyromagnetic factor in the internal magnetic energy term is

$$\left[-\frac{i\hbar\nabla^2}{2M'_n} + 2\left(1+\frac{\sqrt{3}}{2}\right)\left(\frac{e\hbar}{2M'_n c}\right)\frac{\sigma\cdot B}{2}\right]\psi = i\hbar\frac{\partial\psi}{\partial t} \ . \tag{38}$$

The system $(p,e,\bar{\nu})$ is a spin ½ system because the proton spin is opposite to the $(e,\bar{\nu})\equiv e'$ spin. Thus, the neutron and proton forming the deuteron have their spins in the same direction. The fact that the deuteron magnetic moment is the approximate sum of the proton and neutron moments may be explained because the presence of an external magnetic field forces one proton to react by itself while the rest of the system, $p$, $e'$ reacts adjusting its fields as a neutron.

## 6- The many deuteron model.

The presence of orbital angular momentum, so that $\lambda \neq 0$, breaks the stable quasi-static mechanism of the model. The equation determines only one quasi-static system of two protons bound by the strong magnetic interaction, corresponding to states $\nu = \pm\frac{1}{2}$. Nevertheless, the quasi-static proton-electron-proton system discussed supplies another coupling mechanism which allows the combination of more than 2 protons. The $e'$ subsystem strong magnetic moments, of two different proton-electron-proton systems, may align in an anti-parallel formation producing a magnetic pairing of two deuterons. The deuterons loose their separate identity forming a 4-proton excitation.

The alpha particle excitation $\alpha$, may be considered as a 2-deuteron excitation. Since the model is on the equatorial plane, with a $\rho^{-4}$ dependence due to the electronic magnetic field, the resultant equation for the quantum states is still the same Mathieu equation. The only possible Mathieu solution has the same $q_o$ value and only two possible states $\nu = \pm\frac{1}{2}$ to be occupied by protons. This means that the links are only established in proton pairs. The coupling in each pair, due to the presence of the magnetic field of the electrons, is similar to that of the deuteron.

One difference in the model is that now the reduced mass $m$ of the system is approximately half the electron mass $m_e$ due to the presence of two electrons in the system. There is another difference because the addition of the su(2) generators determines that the electromagnetic representation corresponds to a charge 2 system. The corresponding angle between odd an even components is now $\Theta_1^1$ whose tangent is $2^{1/2}$. Instead of eq. (10) we have

$$\left[\nabla^2 + 2m^2\varepsilon^2 + \frac{\mu'^2}{\rho^4}\right]\psi = 0 \ . \tag{39}$$

The effective net magnetic moment $\mu'$ in this equation corresponds to the combination of the two $e'$ subsystems. We expect that it is a fraction $f$ of the original $\mu$. **The required coordinate transformation to obtain Mathieu's equation** is then different from eq. (20),

$$\rho^2 = \frac{\mu'}{\sqrt{2}m\varepsilon}z^2 = \frac{\gamma'}{\sqrt{2m_e m}\varepsilon}z^2 = \frac{\gamma'}{2\sqrt{2}m^2\varepsilon}z^2 \ , \tag{40}$$

which changes $q$ proportionally and determines the energy eigenvalues $E'$ per pair,

$$E'_0 = \varepsilon^2 m = \left(\frac{\sqrt{2}q}{\gamma'}\right)^2 m = -\frac{4}{f^2}q_0^2 m_e = \frac{4}{f^2}E_0 \ . \tag{41}$$

The protons should symmetrically rearrange at the vertices of a square in the equatorial plane. The influence of the two $e'$ subsystems, due to their location on opposite sides and anti-parallel directions, reduces the net moment $\mu'$ by an amount inversely proportional to their squared distances relative to a proton on a square vertex,

$$\frac{|\mu_2|}{|\mu_1|} = \left(\frac{\rho_1}{\rho_2}\right)^2 = \frac{1/4}{5/4} = 0.2 \ . \tag{42}$$

and the fraction $f$ is 0.80.

The binding energy of the system of 2 proton pairs is determined for each proton pair,

$$U' = 2\times 2E'_0 = (8/f^2)U_d = -27.6 \text{ Mev.} \approx -28.3 \text{ Mev.} = U_\alpha \ . \tag{43}$$

For higher proton numbers the system would adjust to the most favorable energy. Therefore, we are lead by the model to consider that the proton-electron-proton links or deuterons act as the essential components of more complex systems in electromagnetic interaction with additional nucleons. This is consistent with the protonic and neutronic numbers of the nuclides. It may also be possible to consider the isobar nuclides $^3H$ and $^3He$ in the same manner.

## 7- Conclusion.

The numbers obtained for the binding energies are surprisingly close to the experimental values, for this crude model. These values admit correction factors due to the approximations made.

From a theoretical point of view it is interesting to know that this "strong" $SU(2)_Q$ electromagnetism, without the help of any other force, generates sufficient attractive short range potentials to provide the binding energy of light nuclides, composed of protons and electrons. Furthermore, it might be that no other nuclear force is required to explain all the properties of nuclides. In any case, implications of the existence of this "strong" electromagnetic SU(2) structure should play a fundamental role in the photo-nuclear interactions, the structure of nuclides, nuclear fusion, the development of nuclear models, gamma ray spectroscopy and other related fields.

In particular, if the binding energy of the deuteron is pure magnetic, its standard fission reaction into a proton and a neutron would be a magnetic multipolar transition of type M1. This transition should be produced by a gamma electromagnetic radiation of type M1. This requires the absorption of the corresponding quantum or M1 gamma photon. In the inverse reaction of fusion, the deuteron should emit a resonant M1 gamma photon in the direction of the deuteron spin, circularly polarized with helicity 1.

From a practical point of view these results should not be overlooked because the implications it might have in the understanding of the process of nuclear fusion of light nuclei. In particular, it may provide possible solutions to the present problems in the development of clean fusion energy.

In principle, apart from technical difficulties, strong localized magnetic fields related to ***directed*** gamma radiation in a nuclear plasma may cause resonance effects which stimulate the same gamma emitting fusion reaction. This reaction in a sector of a magnetically polarized proton-electron-neutron plasma may be amplified by stimulation due to the presence of resonant M1 gamma radiation emitted by another plasma sector with the appropriate arrangement and conditions. In other words, fusion and radiation amplification by stimulation by gamma radiation. The electromagnetic field would then be an active element of a fusion generator rather than only the passive container of a plasma reaction process.

It may be argued, that if this model is realized in nature, there should be additional experimental evidence in its favor. The high magnetic fields in neutron stars may be one evidence. In this respect, we should point out that the most remarkable feature of the electromagnetic SU(2) is its triple geometric structure. In particular this structure should reveal itself as resonance triplets associated to su(2) subexcitations inside nucleons. Experimental evidence for these phenomena is masqueraded as quarks. Rather than fundamental blocks of matter, quarks may be considered excited internal states that actually support the geometric electromagnetic SU(2) interaction.

## Appendix

The odd subspace of $su(2)_Q$, spanned by the two compact odd electromagnetic generators, is isomorphic to the odd subspace of the quaternion algebra, spanned by its orthonormal subset $q^a$. Associated to this orthonormal subset we have the Dirac operator $q \cdot \nabla$ on a curved bidimensional space. This Dirac operator represents the rotation operator $L^2$ on the vector functions on the sphere which also corresponds to the laplacian bidimensional component. We obtain for this action, if we separate the wave function $\varphi$ into the $SU(2)_S$ eigenvector $\phi$ and the $SU(2)_Q$ eigenvector $\psi$, and use the fact that the Levi-Civita connection is symmetric,

$$\left(q^a \nabla_a\right)^2 \psi = q^a q^b \nabla_a \nabla_b \psi = q^{(a} q^{b)} \nabla_a \nabla_b \psi + q^{[a} q^{b]} \nabla_a \nabla_b \psi$$
$$= -g^{ab} \nabla_a \nabla_b \psi = -\Delta \psi = L^2 \psi \ . \tag{44}$$

This equation shows that the squared curved Dirac operator is the Laplace-Beltrami or Casimir operator. Therefore, we may define the odd electromagnetic generator $^-E$ as the quaternion differential operator

$$^-E \equiv q^a \nabla_a = \sqrt{-\Delta} = \sqrt{C^2} \ . \tag{45}$$

The $^-E$ direction in the odd tangent plane is indeterminable because there are no odd eigenvectors common with the even electromagnetic generator $^+E$ and the Casimir generator $E^2$. Nevertheless the absolute value of this quaternion must be the square root of the absolute value of the Casimir quaternion. The absolute values of $^+E$ and $^-E$ define a polar angle $\Theta$ in the su(2) algebra as indicated in the figure.

The angle $\Theta$ is a property of the algebra representations, independent of the normalization as may be verified by substituting $E$ by $NE$. The generators have twice the magnitude of the standard spin generators. This normalization introduces factor of 2 in the respective commutation relations structure constants and determines that the the generator eigenvalues, characterized by the charge quantum numbers $c$, $n$, are twice the standard eigenvalues characterized by the spin quantum numbers $j$, $m$. Nevertheless, it appears convenient to use the two different normalizations for the $SU(2)_Q$ and $SU(2)_S$ isomorphic subgroups in accordance with the physical interpretation of the integer charge and half-integer spin quanta. With the standard normalization, in the base of the common eigenvectors of $^+E$ and $E^2$, we have,

$$^-E^2 |j,m\rangle = j(j+1)|j,m\rangle , \tag{46}$$

$$|^-E||j,m\rangle = |q^a \nabla_a||j,m\rangle = \sqrt{j(j+1)}|j,n\rangle , \tag{47}$$

$$^+E|j,m\rangle = m|j,m\rangle . \tag{48}$$

The electromagnetic generator has an indefinite azimuthal direction but a quantized polar direction determined by the possible translation values. Therefore we obtain, since the absolute values of the quaternions are the respective eigenvalues,

$$\frac{|^-E|}{|^+E|} = \frac{|E^2|^{1/2}}{|^+E|} = \frac{\sqrt{j(j+1)}}{m} \equiv \tan \Theta_m^j = \frac{\sqrt{c(c+2)}}{n} \equiv \tan \Theta_n^c . \tag{49}$$

The internal direction of the potential $A$ and the current $J$ must be along the possible directions of the electromagnetic generator $E$ in $su(2)_Q$. The $A$ components must be proportional to the possible even and odd translations. In consequence the total $A$ vector must lie in a cone, which we call the electrocone, defined by a quantized polar angle $\Theta_n^c$ relative to an axis along the even direction and an arbitrary azimuthal angle.

The complex charged raise and lower generators $^-E^\pm$, which are different from the odd generator $^-E$, are defined in terms de the real generators

$$^-E^\pm = E^1 \pm iE^2 \tag{50}$$

and obey the relations

$$^+E\ ^-E_\pm |c,n\rangle = (n \pm 2)\ ^-E_\pm |c,n\rangle . \tag{51}$$

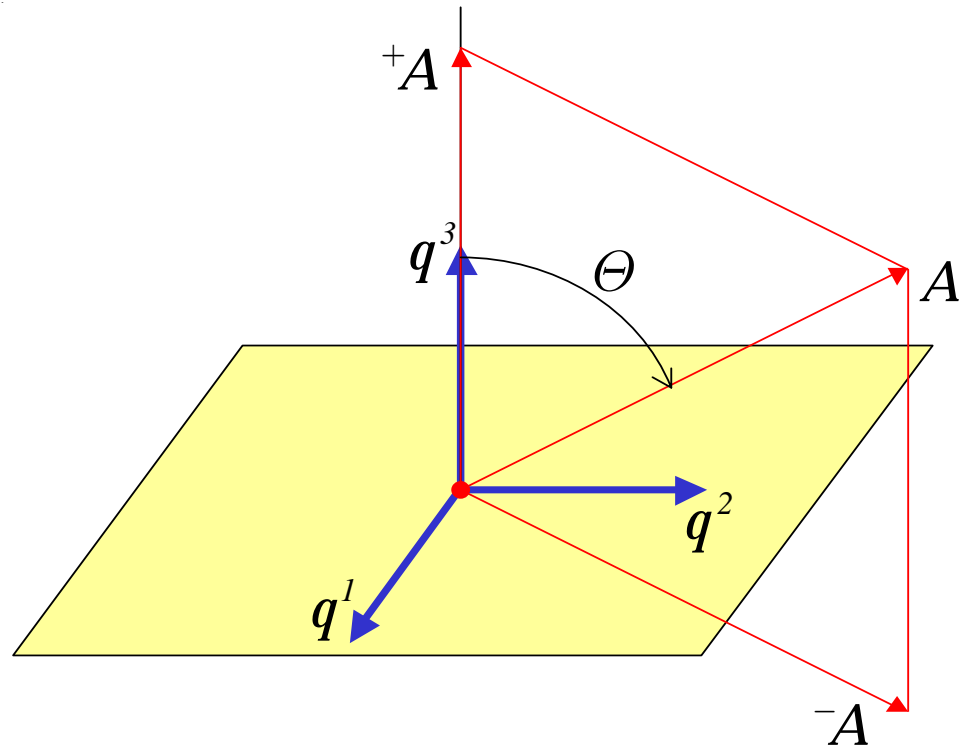

Polar angle $\Theta$ in the $su(2)_Q$ algebra determined by the even and odd generators